\documentclass[twocolumn,prl,aps,floatfix]{revtex4}
\usepackage{amsmath,amssymb}
\usepackage{graphicx,graphics}
\usepackage{fancyhdr}

\begin{document}

\unitlength=1mm

\title{Strong coupling theory for the Jaynes-Cummings-Hubbard model}
\author{S.\ Schmidt}
\author{G.\ Blatter}
\affiliation{Institute for Theoretical Physics, ETH Zurich, 8093 Zurich, Switzerland}

\date{\today}

\begin{abstract}

We present an analytic strong-coupling approach to the phase diagram and elementary excitations of the Jaynes-Cummings-Hubbard model
describing a superfluid-insulator transition of polaritons in an array of coupled QED cavities. In the Mott phase, we find four modes corresponding to particle/hole excitations with lower and upper polaritons, respectively. 
Simple formulas are derived for the dispersion and spectral weights within a strong-coupling random-phase approximation (RPA).
The phase boundary is calculated beyond RPA by including the leading correction due to quantum fluctuations.

\end{abstract}


\maketitle

The recent experimental success in engineering strong interactions between photons and atoms in high-quality
micro-cavities opens up the possibility to use light-matter systems as quantum simulators for many-body
physics~\cite{HB08}. A prominent example is the superfluid-insulator transition of polaritons in an array of coupled QED cavities as
described by the Jaynes-Cummings-Hubbard model (JCHM)~\cite{GT06,HB06}. The competition between strong atom-photon coupling, giving rise to an 
effective photon repulsion (localization), and the photon hopping between cavities (delocalization) leads to a quantum phase
diagram featuring Mott lobes \cite{GT06} reminiscent of those of ultracold atoms in optical lattices as described by the seminal Bose-Hubbard model (BHM)~\cite{FW89}. The JCHM can be implemented, e.g. using single atoms~\cite{Bi05}, excitons~\cite{HB07}, or Cooper pairs~\cite{FG07}. The striking advantage of using coupled micro-cavities, with respect to their optical lattice counterparts, is their individual accessibility.

Although the coupling between QED cavities has not yet been implemented experimentally, the JCHM stimulated exciting theoretical 
work over the last three years, suggesting that polariton systems can be used to simulate various strongly correlated and exotic phases~\cite{AS07,JX07,BK08,CG08,BH09,TG09}.
While such exploratory work is justified in its own right, still very little is known about the fundamental excitations of the JCHM.
The phase boundary of the superfluid-insulator transition has been calculated using mean-field decoupling~\cite{GT06,AS07}, Monte-Carlo~\cite{ZS08,PE09} and 
variational cluster approaches~\cite{RF07,AH08} in two and three dimensions. Only two papers have explored the fundamental excitations of the system~\cite{AH08,PE09}.
All of these results, even on a mean-field level, rely on more or less heavy numerical computation due to the intricate composite nature of polaritons.
To the best of our knowledge, no analytic results are available neither for the phase boundary nor for the excitations of the JCHM.

In this paper, we show that a linked-cluster expansion pioneered for the Fermi-Hubbard model (FHM)~\cite{Me91} and recently applied to the BHM~\cite{OP08} can be used to obtain simple, analytic formulas for the phase diagram as well as excitation spectra for arbitrary temperatures, detuning parameter and lattice geometries. We find two new modes, which have been overlooked in previous numerical approaches and discuss dispersion and spectral weights within strong-coupling RPA. Furthermore, we study the effect of quantum fluctuations on the phase boundary of the superfluid-insulator transition.
In two dimensions, our results agree well with recent Monte-Carlo calculations. In three dimensions, where numerical
data is currently not available, we present a quantitatively accurate calculation of the quantum phase diagram.

The Hamiltonian of the JCHM is given by
\begin{equation}
\label{jchm}
H=\sum_i h^{\rm JC}_i - \mu N - J \sum_{\langle i j \rangle} a^\dagger_i a_j\,,
\vspace{-0.1cm}
\end{equation}
where $h^{\rm JC}_i$ denotes the local Jaynes-Cummings Hamiltonian 
$h^{\rm JC}_i = \omega_c\, a^\dagger_i a_i + \omega_x \sigma_{i}^+\sigma_{i}^-  +  g (\sigma_{i}^+ a_i +\sigma_{i}^- a^\dagger_i)$
with cavity index $i$, photon creation (annihilation) operators $a_i^{(\dagger)}$ and atomic raising (lowering) operators $\sigma_i^{+(-)}$. 
The cavity mode frequency is $\omega_c$, the two atomic levels are separated by the energy $\omega_x$ and the atom-photon coupling is given by $g$ (we set $\hbar=1$).  The total number of excitations, i.e., polaritons $N=\sum_i ( a^\dagger_i a_i + \sigma_{i}^+\sigma_{i}^-)$, is conserved and fixed by the chemical potential $\mu$.  The third term in (\ref{jchm}) describes the delocalization of photons over the whole lattice due to hopping between nearest neighbour cavities with amplitude $J$. It competes with an effective on-site repulsion between photons mediated by the atom-photon coupling. This competition leads to Mott lobes in the quantum phase diagram \cite{GT06}.

A rough estimate for the size of these Mott lobes can be obtained by calculating perturbatively the cost of adding or removing a polariton. In the atomic limit ($J=0$) the eigenstates of the Hamiltonian (\ref{jchm}) are the dressed polariton states labelled by the polariton number $n$ and upper/lower branch index $\sigma=\pm$. For $n>0$ they can be written as a superposition of  a Fock state with $n$ photons plus atomic ground state $|n, g\rangle$ and $(n-1)$ photons with the atom in its excited state  $|(n-1), e\rangle$,
\begin{eqnarray}
|n +\rangle &=& \sin\theta_n |n\,,g\rangle + \cos\theta_n |(n-1)\,,e\rangle\,, \nonumber\\
|n -\rangle &=& \cos\theta_n |n\,,g\rangle -  \sin\theta_n |(n-1)\,,e\rangle\,,
\end{eqnarray}
with the angle $\tan \theta_n=2 g \sqrt{n}/(\delta+2\chi_n)$, $\chi_n=\sqrt{g^2 n + \delta^2/4}$ and the detuning parameter $\delta=\omega_c-\omega_x$. The corresponding eigenvalues are
\begin{equation}
\epsilon_n^{\sigma}=-(\mu-\omega_c) n - \delta/2 +\sigma\,\chi_n\,,\quad \sigma=\pm\,.
\end{equation}
The zero polariton state $|0-\rangle=|0\,,g\rangle$ is a special case with $\epsilon_0^-=0$. 
Upper and lower polariton energies are separated by the Rabi splitting $\Omega_n=2\chi_n$. 
For the calculation of the quantum phase diagram at small tunneling $J\ll g$ we can neglect the upper polariton branch.
In a straightforward first-order degenerate perturbation theory in $J$, the chemical potentials at which the addition/removal of a lower polariton
costs no energy is given by (we assume a hypercubic lattice in $D$ dimensions)
\begin{eqnarray}
\label{chem}
\mu_{p}-\omega_c&=&\chi_n - \chi_{n+1} - 2 D J (f^{--}_{n+1})^2 +{\mathcal{O}}(J^2/g)\,,\nonumber\\
\mu_{h}-\omega_c&=&\chi_{n-1} - \chi_{n} + 2 D J (f^{--}_{n})^2+{\mathcal{O}}(J^2/g)\,;
\end{eqnarray}
the matrix elements $f^{\sigma \nu}_n= \langle n\,\sigma | a^\dagger | (n-1)\, \nu\rangle$ can be easily expressed in terms of the angle $\theta_n$ and are given by
$f^{\sigma \nu}_n=\left( \sqrt{n}+\sigma\,\nu\,\sqrt{n-1}\right)/2$ for $n>1$  ($f^{\sigma -}_1=1/\sqrt{2}$) at zero detuning ($\delta=0$).
The two equations in (\ref{chem}) define the upper and lower phase boundary in the quantum phase diagram in Fig.~\ref{fig1} for small values of the hopping parameter $J/g$. The point where the two lines meet (i.e., $\mu_p=\mu_h$) is $J_c\sim 0.1 g$ for $n=1$ and represents an upper limit for the size of the first Mott lobe. 
By going to higher order in the perturbative expansion for the ground-state energy and subsequent resummation of the strong-coupling series, one could in principle determine the exact location of the phase boundary. This has recently been achieved for the BHM \cite{SP09,TH09}. 

In this paper, we will follow a different approach and study directly the photonic Matsubara Green's function 
$ G_{ij}(\tau;\tau')= - \langle {\mathcal{T}} a_i(\tau) \bar{a}_j(\tau')\rangle$
with the time-ordering operator {${\cal{T}}$ and the Heisenberg operator $\bar{a}_j(\tau')=e^{H\tau'}a_j^\dagger e^{-H\tau'}$. A suitable method for the evaluation of the Matsubara Green's function is a linked-cluster expansion in terms of local cumulants originally developed by Metzner {\it et al.}~\cite{Me91} for the FHM. Each term of the linked-cluster expansion can be written diagrammatically in terms of n-particle cumulants represented by 2n-leg vertices and tunneling matrix elements symbolized by propagating lines connecting two vertices. The strong-coupling expansion provided by the linked-cluster method is applicable to the JCHM because (i) the atomic limit Hamiltonian is local, (ii) anomalous averages of the photon operator with respect to the eigenstates of the local Hamiltonian vanish, i.e.,
$\langle n \sigma | (a^\dagger)^k | n \sigma \rangle = 0\quad\mbox{for}\quad k \in N$ (since a single photon excitation always changes the polariton number).

In the atomic limit, the Green's function is given by $G_{0 ij}(\tau;\tau')=G_{0 i}(\tau;\tau')\,\delta_{ij}= - \langle {\mathcal{T}} a_i(\tau) \bar{a}_i(\tau')\rangle_0\,\delta_{ij}$, where the average $\langle\dots\rangle_0$ is taken with respect to the eigenstates of the local Hamiltonian, i.e., the first two terms in (\ref{jchm}). For a spatially
homogeneous system we can drop the site index $i$ and straightforwardly calculate $G_0(\tau;\tau')$.  After a Fourier transformation we obtain 
\begin{equation}
\label{cum1}
G_0(\omega_m)\hspace{-0.05cm}=\hspace{-0.15cm}\sum_{n,\sigma,\mu}\hspace{-0.05cm} \frac{e^{-\beta\epsilon^\sigma_n}}{Z}\left( \frac{z_{n+1}^{\mu\sigma}}{\Delta_{n+1}^{\mu\sigma} - i \omega_m}-\frac{z_{n}^{\sigma\mu}}{\Delta_{n}^{\sigma\mu} - i \omega_m} \right)
\end{equation}
with the partition function $Z=\sum_{n\sigma}e^{-\beta\epsilon^\sigma_n}$ and bosonic Matsubara frequencies $\omega_m=2\pi m/\beta$  ($\beta=1/(k_{\scriptscriptstyle B} T)$ with temperature $T$ and Boltzmann constant $k_{\scriptscriptstyle B}$). In (\ref{cum1}) we defined $z^{\mu\sigma}_n=(f_{n}^{\mu\sigma})^2$ and $\Delta^{\mu\sigma}_n=\epsilon_{n}^\mu-\epsilon_{n-1}^\sigma$.
\begin{figure}
\includegraphics[scale=0.41]{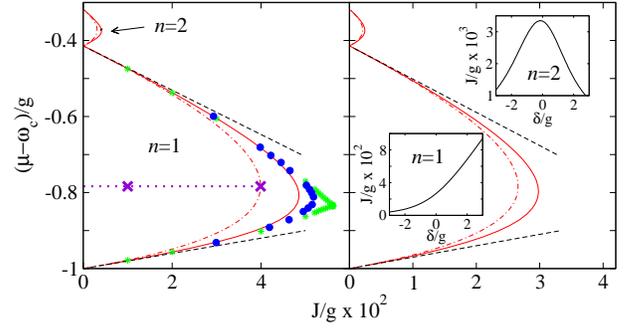}
\vspace{-0.3cm}
\caption{\label{fig1} Quantum phase diagram for a hypercubic lattice in $D=2$ (left figure) and $D=3$ (right figure). We show two Mott lobes for $n=1,2$ at zero detuning $\delta=0$.  We compare first-order perturbation theory (dashed), RPA (dot-dashed) and quantum fluctuations (solid) with recent results from a quantum Monte-Carlo (filled dots) \cite{ZS08} and variational cluster (stars) \cite{AH08} approach. The two crosses (left figure) connected by the dotted line mark the parameter values used in Fig.~\ref{fig2}. The insets show the critical hopping strength $J_c/g$ at the tip of the lobe as a function of the detuning $\delta$ for $n=1$ and $n=2$ calculated within strong-coupling RPA in $D=3$ dimensions. }
\vspace{-0.5cm}
\end{figure}
We sum an infinite set of diagrams by calculating the irreducible part of the Green's function $K({\bf k},\omega_m)$ which is connected to the full Green's function via the equation $G({\bf k},\omega_m)=K({\bf k},\omega_m)/[1-J({\bf k}) K({\bf k},\omega_m)]$ with the lattice dispersion $J({\bf k})=2J\sum_{i=1}^D \cos{{\bf k \cdot a_i}}$ (${\bf a}_i$ denotes a lattice vector). To second order in $J$ we get 
\vspace{-0.3cm}
\begin{eqnarray}
\label{irred}
\jot10mm
K(\omega_m)&=&\raisebox{-0mm}{\includegraphics[width=14mm]{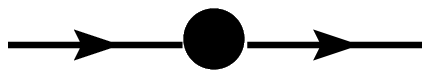}}\,\,\,+\,\,\, \raisebox{-0mm}{\includegraphics[width=14mm]{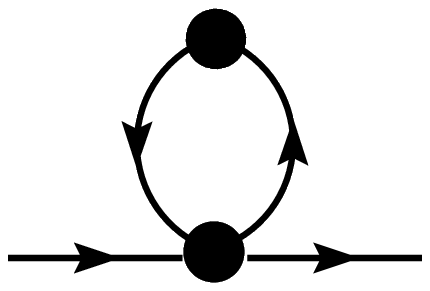}}\nonumber\\
&=& G_0(\omega_m) + 2 D J^2\,Q(\omega_m)
\end{eqnarray}
with 
$Q(\omega_m)\hspace{-0.05cm}=\hspace{-0.15cm}\int_0^\beta \hspace{-0.15cm}d\tau  d\tau_1  d\tau_2\, C^{(2)}(\tau_1 0;\tau_2 \tau) G_0(\tau_2;\tau_1) e^{i\omega_m\tau}$.
The quantum fluctuation correction $Q(\omega_m)$ involves the two-particle cumulant $C^{(2)}(\tau_1 0;\tau_2 \tau)$, which is related to the local two-particle Green's function $G^{(2)}_0(\tau_1 0;\tau_2 \tau)=\langle {\mathcal{T}} a_i(\tau_1) a_i(0) \bar{a}_i(\tau_2) \bar{a}_i(\tau)\rangle_0$ via $C^{(2)}(\tau_1 0;\tau_2 \tau)=G^{(2)}_0(\tau_1 0;\tau_2 \tau)-G_0(\tau_1;\tau_2 )G_0(0;\tau)-G_0(\tau_1;\tau )G_0(0;\tau_2)$. 
The algebraic expressions for the two-particle cumulant and the quantum correction $Q(\omega_n)$ are lengthy and will be given elsewhere \cite{SB}. 

The inverse Green's function tells us immediately about the phase boundary $G^{-1}({\mathbf 0},0)|_{J_c(\mu)}=0$ and the dispersion relation $G^{-1}({\mathbf k},i\omega_m\rightarrow \omega+i0^+)=0$.
We introduce the strong-coupling self-energy $\Sigma({\bf k},\omega_m)$ via $G({\bf k},\omega_m)^{-1}=G_0(\omega_m)^{-1}-\Sigma({\bf k}, \omega_m)$ and obtain from (\ref{irred}) to second order
$\Sigma({\bf k},\omega_m)=J({\bf k}) + 2 D J^2 Q (\omega_m)/G_0(\omega_m)$.
The first term on the r.h.s. is usually called the strong-coupling random-phase approximation (RPA)~\cite{SD05}, whereas the second term denotes the leading correction due to quantum fluctuations. The RPA corresponds to a summation of all self-avoiding walks (chain diagrams) through the lattice. The leading quantum correction includes in addition all one-time forward/backward hopping processes (bubble diagrams) between two neighbored sites. 
Although we have carried out our calculations at finite temperature, we will only consider the $T=0$ case from now on. The effect of thermal fluctuations will be studied elsewhere \cite{SB}.
\begin{figure}[t]
\includegraphics[scale=0.4]{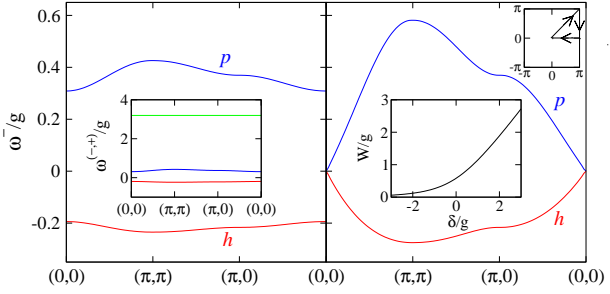}
\vspace{-0.2cm}
\caption{\label{fig2} Particle/hole dispersion of the conventional modes $\omega^-_{(p,h)}$ for $D=2$ and $n=1$ at zero detuning $\delta=0$. Left figure: Deep inside the Mott insulator (see cross in Fig.~\ref{fig1}) for $(\mu-\omega_c)/g=-0.78$ and $J/g=0.01$. The inset shows the conventional modes $\omega^-_{(p,h)}$ together with the conversion mode $\omega^+_{p}$. Right figure:  At the phase boundary (see cross in Fig.~\ref{fig1}) for  $(\mu-\omega_c)/g=-0.78$ and $J/g=0.04$. The inset shows the bandwidth $W$ of the conventional particle mode $\omega^-_p$ as a function of detuning $\delta$ at the tip of the lobe.}
\vspace{-0.5cm}
\end{figure}

At the quantum phase transition the energy of long wavelength fluctuations vanishes and we get an explicit expression for the 
critical hopping strength
\begin{equation}
\label{pd}
J_c^{-1}=D\,G_0(0)\big( 1 + \sqrt{1 + 2 Q(0)/(D\, G_0^3(0))}\big)\,.
\end{equation}
If we ignore the second term under the square root in (\ref{pd}) we obtain the RPA phase boundary 
$1/J_c=2D\,G_0(0)$ shown as a dashed-dotted line in Fig.~\ref{fig1}. We have checked that 
this analytic result agrees exactly with the phase boundary obtained from a numerical decoupling mean-field approach 
as presented in~\cite{GT06,AS07}. We can go beyond this result by including quantum fluctuations in (\ref{pd}) leading to an improved phase boundary (solid line in Fig.~\ref{fig1}). In two dimensions our result agrees well with recent Monte-Carlo calculations \cite{ZS08}  and confirms the smoothness of the lobes found in \cite{ZS08} (different from \cite{AH08}, where a variational cluster approach has been used). There is still a small deviation near the tip of the lobe, where quantum corrections are most important. In the right panel of Fig.~\ref{fig1} we present quantitatively accurate results for the phase diagram in $D=3$. Note, that the strong-coupling expansion becomes more accurate in higher dimensions. 
\begin{figure}
\includegraphics[scale=0.4]{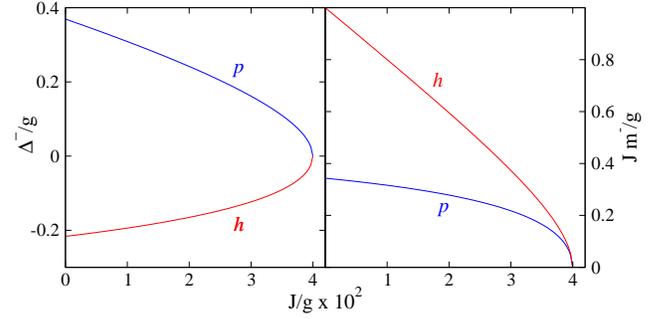}
\vspace{-0.4cm}
\caption{\label{fig3} Particle/hole gaps $\Delta^-_{(p,h)}$ (left figure) and effective masses $m^-_{(p,h)}$ (right figure) of the conventional modes for $D=2$ and $n=1$ at zero detuning $\delta=0$ as a function of the tunneling strength $J/g$ for $(\mu-\omega_c)/g=-0.78$ (along the dotted line in Fig.~\ref{fig1}).}
\vspace{-0.5cm}
\end{figure}

We now turn to the discussion of excitations, which we calculate within RPA. An analytic continuation of the Matsubara Green's function via $i\omega_n\rightarrow \omega + i 0^+$ yields the retarded real-time Green's function. Its poles and residues yield the dispersion relations and mode strengths of the fundamental excitations. In general, we obtain four poles, the conventional (Bose-Hubbard like) lower polariton particle (hole) modes $\omega^-_{(p,h)}$ and two modes $\omega^+_{(p,h)}$ which correspond to an upper polariton particle (hole) excitation (conversion modes). The presence of the latter signals a clear deviation from the usual Bose-Hubbard like physics and is due to the composite nature of polaritons. The conversion modes exist already in the atomic limit and were also found in a recent Monte Carlo study \cite{PE09}, but have been overlooked by a variational cluster approach \cite{AH08}. One reason might be that their bandwidth and strengths are very small as compared to the conventional modes. 
We can thus set their dispersion relations $\omega^+_{(p,h)}$ and mode strength $s^+_{(p,h)}$ approximately equal to $\omega^+_{(p,h)}\approx \Delta^+_{(p,h)}\quad\mbox{and}\quad s^+_{(p,h)}\approx z^+_{(p,h)}$
with the atomic-limit particle (hole) gaps $\Delta_p^\sigma\equiv\Delta_{n+1}^{\sigma-}$ ($\Delta_h^\sigma\equiv\Delta_n^{-\sigma}$) and mode strengths $z_p^\sigma\equiv z_{n+1}^{\sigma-}$ ($z_h^\sigma\equiv-z_n^{-\sigma}$), respectively. 
If we neglect their contribution to the one-particle cumulant (\ref{cum1}), we can derive simple analytic formulas for the dispersion relations of the conventional modes
\begin{equation}
\label{dispan}
\omega^-_{(p,h)}=\left( \Delta_+\, -\, J({\bf k})\, z_+\, \pm\, \Omega \right)/2
\end{equation}
with $\Omega = \sqrt{ \Delta_-^2 + J({\bf k})^2\,z_+^2 - 2  J({\bf k})\, z_-\, \Delta_-}$
and the abbreviations $\Delta_\pm=\Delta^-_{p}\pm \Delta^-_{h}$ and $z_\pm=z^-_{p}\pm z^-_{h}$.
The strength of the modes are given by
\begin{equation}
\label{weightsan}
s^-_{(p,h)}=\frac{ z_+\, \omega^-_{(p,h)} - z^-_{(p,h)} \Delta^-_{(h,p)}  - z^-_{(h,p)} \Delta^-_{(p,h)} }{\omega^-_{(p,h)}-\omega^-_{(h,p)}}\,.
\end{equation}
The dispersions in (\ref{dispan}) are plotted in Fig.~\ref{fig2} deep inside the Mott regime and at the tip of the lobe with $n=1$. 
The energy needed for a conventional excitation
is an order of magnitude smaller than for a conversion excitation.
The strengths of the conventional modes $s^-_{(p,h)}$ grow with increasing tunneling strength (and diverge at the tip of the lobe similar to what is found for the BHM \cite{MT08}) while $s^+_{(p,h)}$ stays approximately constant. If we are only interested in low energy excitations we can thus indeed neglect contributions from upper polaritons. 

Deep inside the Mott insulator particle and hole excitations are gapped. If the phase boundary is approached away from the tip of the lobe, either the particle (upper phase boundary) or the hole (lower phase boundary) gap vanishes linearily  $\Delta\sim |J-J_c|$. At the phase boundary the dispersion relation remains quadratic $\omega\sim k^2$. The situation changes at the tip of the lobe, where particle and hole gaps vanish simultaneously with a square-root behavior $\Delta\sim |J-J_c|^{1/2}$, while their dispersions become linear $\omega\sim k$ (see Fig.~\ref{fig2}). This indicates a special transition at the tip of the lobe, where also the effective masses vanish (see Fig.~\ref{fig3}), reminiscent of an emergent particle-hole symmetry. In the opposite limit $J\rightarrow 0$ the effective masses approach infinity as $m^-_{(p,h)}=\pm 1/(2 \,J\, z^-_{(p,h)})$.
The behavior of the dispersion at the critical point $J=J_c$ in the long-wavelength limit ${\bf k}\rightarrow 0$ determines the dynamical critical exponent $z$ defined by $\omega \sim \xi^{-z}\sim k^z$ with the diverging correlation length $\xi\sim|J-J_c|^{-\nu}$ and its associated critical exponent $\nu$. From the discussion above, we conclude that the dynamical critical exponent has the generic value $z=2$ everywhere in the phase diagram except for the special multi-critical point at the tip of the lobe where it changes to $z=1$. Very recently this result has been confirmed using an effective action approach \cite{KH09}. At ${\bf k=0}$ the gap vanishes as $\Delta\sim |J-J_c|^{z\nu}$ when the tunneling strength approaches its critical value $J_c$. Thus we have $\nu=1/2$ everywhere in the phase diagram. We conclude that at least on a mean-field RPA level, the JCHM has the same critical exponents as the BHM \cite{FW89} and is thus in the same universality class. 
The results in (\ref{dispan}) and (\ref{weightsan}) can be seen as a generalization of the expressions for the dispersion relations and mode strengths of the BHM~\cite{OS01}. The usual BHM physics is retrieved if one ignores the upper polariton mode in (\ref{cum1}) and sets $f^\sigma_n\equiv f_n=\sqrt{n}$ and $\epsilon^\sigma_n\equiv \epsilon_n=(U/2)n(n-1)-\mu n$. 

The physics of the JCHM is richer due to the presence of the experimentally important detuning parameter $\delta$. Indeed, detuning from the resonance $\delta=0$ allows to fine-tune the effective repulsion between photons and drive the system from the Mott into the superfluid state \cite{GT06, ZS08}.
In Fig.~\ref{fig1} we show that the critical hopping strength decreases as a function of the detuning parameter $\delta$ independent of its sign for $n=2$. This is generally true for Mott lobes with more than one polariton $n>1$. However, the $n=1$ Mott lobe becomes smaller for negative and larger for positive detuning. This distinct behavior is due the special nature of hole excitations from the $n=1$ polariton state to the zero-polariton state, which has an energy $\epsilon_0^-=0$ independent of detuning. 

At the end we comment on the experimental verifiability of our results. A setup based on quantum dot excitons embedded in a photonic crystal was discussed in~\cite{NU08}. After proper initialization of a Mott insulator ground-state, the dispersion relations of the excitations could be measured using transmission spectroscopy~\cite{WC07}

\begin{acknowledgments}
We thank H. Tureci, A. Pelster, I. Carusotto, A. Badolato, T. Esslinger, and A. Imamoglu for discussions and acknowledge financial support from the Swiss National Foundation through the NCCR MaNEP. 
\end{acknowledgments}

\vfill
\end{document}